\documentclass[%
 reprint,
 amsmath,amssymb,
 aps,
]{revtex4-1}

\usepackage{graphicx}
\usepackage{dcolumn}
\usepackage{bm}
\usepackage{gensymb}

\begin{document}

\preprint{APS/123-QED}

\title{Nanopatterned electron beams for temporal coherence and deterministic phase control of x-ray free-electron lasers}

\author{W.S. Graves}
\email{wsg@asu.edu}
\author{S.L.Y. Chang}%
\author{C. Dwyer}%
\author{P. Fromme}%
\author{M. Holl}%
\author{B.D.A. Levin}%
\author{L.E. Malin}%
\author{M.A. Roldan}%
\author{J.L. Vincent}%
\author{J.C.H. Spence}%
 \affiliation{%
 Arizona State University, Tempe, Arizona 85287, USA
}%
\author{E.A. Nanni}%
\author{R.K. Li}%
\author{X. Shen}%
\author{S. Weathersby}%
 \affiliation{%
	SLAC National Accelerator Laboratory, Menlo Park, CA 94025 USA
}%
\author{A. Sandhu}%
\affiliation{%
	Department of Physics, University of Arizona, Tucson, Arizona 85721 USA
}%




\date{\today}

\begin{abstract}
We demonstrate the ability to create electron beams with high-contrast, nanometer-scale density modulations as a first step toward developing full control of the phase fronts of an x-ray free-electron laser.  The nanopatterned electron beams are produced by diffracting electrons through thin single-crystal silicon membranes that are lithographically patterned to different thicknesses.  For transform-limited x-ray production the desired pattern is a series of regularly spaced lines (i.e. a grating) that generate uniformly spaced nanobunches of electrons, however nearly any pattern can be etched in the silicon, such as frequency-chirps or multiple patterns of different ‘colors’ or line spacings.  When these patterns are transferred from the spatial to the temporal dimension by accelerator electromagnetic optics they will control the phase fronts of coherent x-rays, giving unprecedented deterministic control over the phase of ultrashort x-ray pulses.  In short, this method allows the time-structure for a fully coherent x-ray beam to be generated from a pattern written on a semiconductor wafer by lithography.

\end{abstract}

\maketitle


X-ray free-electron lasers (XFELs) produce intense, transversely coherent, ultrashort x-ray pulses containing enough photons to produce a diffraction pattern from a nanocrystal in a single shot, while outrunning most effects of radiation damage.  XFELs are proving to be powerful tools across a range of applications including crystal structure and dynamics of biological molecules in their native environment \cite{Spence2017}, fundamental charge and energy dynamics in molecules \cite{Zhang2014}, emergent phenomena in correlated electron systems \cite{Forst2014}, single-particle structure and dynamics \cite{Kurta2017}, and matter in extreme environments \cite{Kraus2017}.  XFEL properties can be further improved by gaining control of the radiation's temporal phase.  Current XFELs are transversely coherent but lack temporal coherence, resulting in noisy fluctuations in spectrum and intensity from shot to shot.  This is due to their reliance on self-amplification of spontaneous emission (SASE) \cite{Bonifacio1984,Kim1986}, that is, the output is amplified shot noise from electrons with random initial phases. 

\begin{figure}
	\centering
	\includegraphics[width=8.6cm]{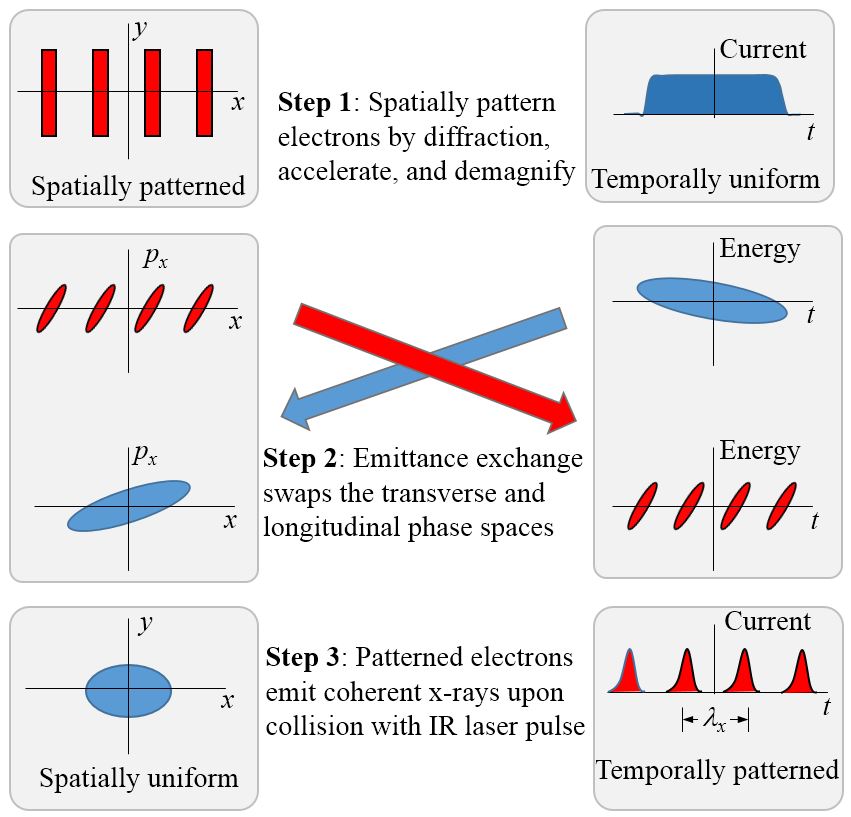}
	\caption{Schematic of the 3 steps necessary to produce coherent x-rays from a nanopatterned electron beam.  The present work provides the first experimental demonstration of Step 1.}
	\label{Fig01}
\end{figure} 

\begin{figure}
	\centering
	\includegraphics[width=8.6cm]{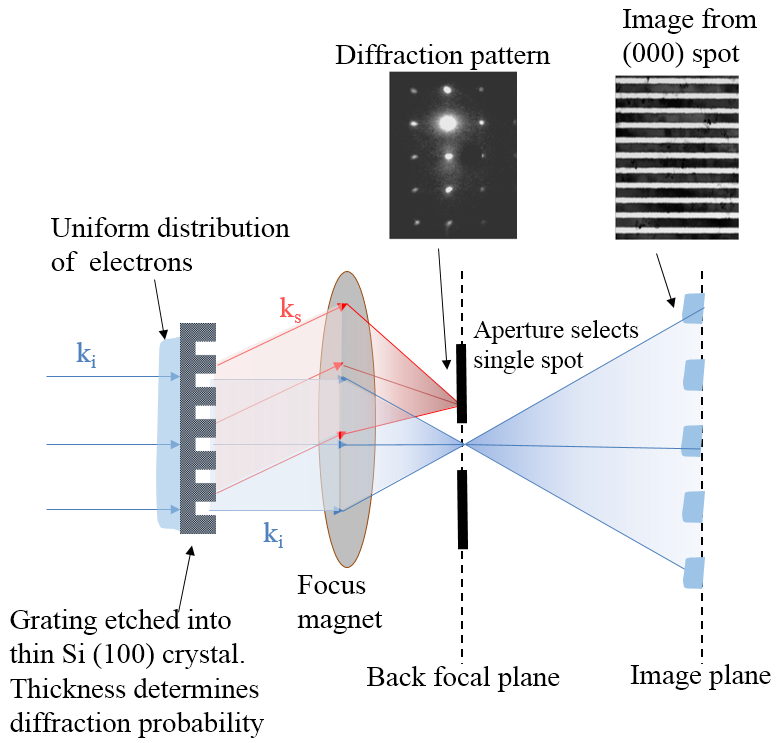}
	\caption{Illustration showing electron diffraction from a thin single crystal Si grating to create a nanopatterned beam.  Diffraction occurs from crystal planes rather than grating structure.  The Si thickness determines fraction of electrons into each Bragg spot.  Spatial pattern is created by blocking all but a single spot. Either a bright field image (shown) or dark field may be generated.}
	\label{Fig02}
\end{figure} 

We recently proposed \cite{Nanni2018, Graves2012} a 3-step method of controlling the x-ray phase by creating nanometer scale density modulations in the electron beam before radiation emission (Figure \ref{Fig01}). In addition to gaining control of the radiation's temporal properties, our concept shrinks the size and cost of an XFEL by orders of magnitude by replacing the magnetic undulator with a powerful picosecond infrared (IR) laser pulse.  The much shorter period of the IR laser compared to a magnetic undulator ($\mu m$ vs $cm$) lowers the electron energy needed to generate x-rays from several GeV to tens of MeV. 

Here we report the successful demonstration of the first step consisting of producing spatially nanopatterned beams via diffraction through thin lithographically patterned Si membranes. Fully coherent x-ray pulses with tailored spectral and temporal properties will open new opportunities in x-ray science.  In chemical and biological systems, charge migration and energy redistribution dynamics involve coherent reaction pathways and coupled electronic and nuclear motions spanning attosecond to picosecond timescales \cite{Worner2017}. In order to map such dynamics using pump-probe schemes, it is essential to have temporally coherent x-ray pulses. In particular, multi-dimensional non-linear x-ray spectroscopy of quantum coherence in molecular systems and correlated materials requires the application of a precisely controlled sequence of x-ray pulses with full spatial and temporal control of the phase fronts \cite{Schweigert2007}. 

The SASE process depends on amplification of initial shot-noise \cite{Wang1984,Saldin2010} in the electron beam with the result that the XFEL radiation has stochastic properties with random phase jumps that produce a spiky single-shot power spectrum with large shot-to-shot intensity fluctuations.  This startup from noise could be overcome if a coherent seed existed that could be amplified.  A variety of seeding techniques have been implemented \cite{Yu2000, Yu2003, Stupakov2009, Xiang2010} but are limited by either the low photon energy of the seed or, in the case of self-seeding \cite{Feldhaus1997, Geloni2010}, from effects of the original shot noise fluctuations. For photon energies above a few hundred eV no coherent seed radiation exists.

\begin{figure}[!htb]
	\centering
	\includegraphics[width=8.6cm]{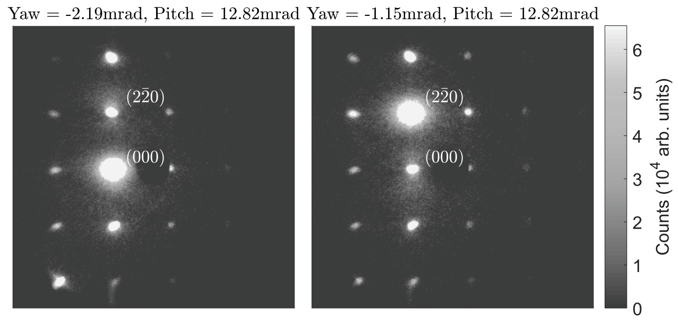}
	\caption{Experimental data from SLAC’s UED facility.  The Si membrane is tilted to different incident angles for each image demonstrating the ability to diffract nearly the entire distribution into different single Bragg spots.  Left shows 80\% of beam forward scattering into (000) spot at the indicated pitch and yaw angles of the sample.  Right shows 80\% of beam scattering into $(2\overline{2}0)$ spot at indicated pitch/yaw.}
	\label{Fig03}
\end{figure} 

In contrast to SASE or seeding, our method arranges the electrons into discrete bunches (nanobunching) with periodicity equal to the desired x-ray wavelength.  The nanobunched  electrons then emit coherently at that wavelength \cite{Graves2012, Graves2017a, Graves2013}.  The first step in this method is to generate the desired electron beam pattern by diffracting relativistic electrons in the transmission geometry through a thin silicon membrane \cite{Malin2019, Malin2017, Zhang2017}.  The membranes are patterned into a grating-like structure with alternating thin and thick sections.  Diffraction from the Si crystal planes determines the fraction of electrons (Figures \ref{Fig02} and \ref{Fig03})elastically scattered into particular Bragg spots. The purpose of the grating structure is to spatially modulate the fraction of electrons diffracted into particular Bragg spots as illustrated in Figure \ref{Fig02}.  We may then create dark- or bright-field images of the membrane, using a single Bragg beam selected by an aperture, showing alternating bright and dark stripes across the electron beam.  At unity magnification these transversely-patterned nanobunches have the same spatial periodicity as the Si structure, but this period may be continuously adjusted over a wide range since the pattern may be demagnified by a factor of 100 or more using magnetic lenses to scale the nanopattern into the x-ray range.

\begin{figure*}[!htb]
	\centering
	\includegraphics[width=17cm]{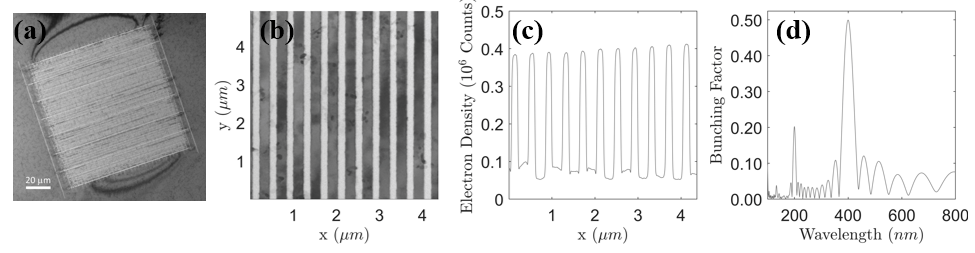}
	\caption{a) Transmission electron micrograph that shows grating area patterned on thin Si membrane;  b) shows bright-field image of Si grating.  Bright areas are high electron density from Si grooves.  Electrons that pass through ridges are diffracted and blocked resulting in dark stripes;  c) shows projected density indicating a 4:1 contrast ratio between electrons transmitted through grooves vs ridges; and d) shows FEL bunching factor (modulus of Fourier transform), similar to saturated FEL.}
	\label{Fig04}
\end{figure*} 

To produce XFEL radiation after generating the patterned electrons, two further steps are needed as shown in Figure \ref{Fig01}.  The second step is to accelerate the nanopatterned beam and use emittance exchange \cite{Cornacchia2002, Sun2010, Carlsten2011, Nanni2015} to swap the pattern from the transverse distribution to the longitudinal or temporal distribution.  The third step is to propagate the beam through an undulator or colliding laser pulse to produce synchrotron radiation, coherent in this case due to the repetitive density modulation, and amplify the coherent signal through the FEL instability.

A simple example of the control this method offers is to change the electron spot size on the membrane by focusing, demonstrated in our experiments and described below, illuminating different numbers of grating periods, thus producing different numbers of nanobunches. This will control both the x-ray pulse length and its bandwidth, given by the reciprocal of the number of grating periods illuminated.  Our experiments demonstrate a high contrast ratio in the electron pattern so that the x-ray output pulse is expected to be nearly transform limited.

A simple example of the control this method offers is to vary the eventual x-ray pulse length and its bandwidth, given by the reciprocal of the number of grating periods illuminated, by simply changing the electron spot size on the membrane using magnetic focusing. Changing the spot size illuminates different numbers of grating periods, producing different numbers of nanobunches.  We demonstrate and discuss this effect below. Many other manipulations are possible, such as multiple spots to produce multiple pulses with precise delays and with different colors if desired.

Our first set of experiments used an unpatterned Norcada Uberflat (100) single-crystal silicon membrane 200nm thick that was studied at ASTA, SLAC’s Ultrafast Electron Diffraction (UED) facility \cite{Weathersby2015, Shen2018}. The ASTA facility uses an RF photoinjector to produce femtosecond electron bunches at a few MeV energy. Our experiments ran at 2.26 MeV with 100 fs bunches over a range of charges from 10 fC to 1 pC.  These experiments were designed to test aspects of the beam patterning concept including whether photoinjectors produce beams of adequate quality to cleanly diffract, whether crystal damage is an issue, and whether we can accurately predict the intensities of Bragg spots at different incident angles for the relativistic beams passing through the thin membranes.  The results were positive on each of these key aspects and are reported in detail in Malin et al \cite{Malin2019}.  An example of the high contrast possible is shown in Figure \ref{Fig03}. After aligning the crystal planes of the membrane to the sample holder, angle scans using simulations for guidance were carried out to simultaneously maximize the $(2\overline{2}0)$ beam and extinguish the (000) beam. Figure \ref{Fig03} compares two diffraction patterns.  On the left side the direct beam is maximized at a pitch angle of -2.19 mrad, and on the right side the $(2\overline{2}0)$ reflection is maximized at a pitch angle of -1.15 mrad.  We achieved approximately 80\% transmission of the beam into the $(2\overline{2}0)$ reflection in the diffraction pattern relative to the incident electron beam in this case. At this pitch angle the (000) beam is reduced to about 5\% of the incident intensity with the remaining few percent scattering into other peaks.  Note that the studies in \cite{Malin2019} indicate that about 20\% of the incident beam is inelastically scattered by electron-plasmon interactions, and that another few percent is absorbed in the crystal.  These effects mean that the most efficient route to patterning is to use the (000) spot for the transmitted beam and to diffract the unwanted electrons into other low order reflections that are then blocked.

Selecting distinct Bragg diffraction peaks requires the beam divergence angle at the membrane to be significantly smaller than the first-order Bragg angle, which is 1.2 mrad at 2.26 MeV. In addition the relative energy spread must be low enough that chromatic aberration in the focusing system does not cause the peaks to overlap. These requirements on divergence and energy spread are met by the photoinjector electron beam as shown in Figure \ref{Fig03}.

The lifetime of the silicon membrane will eventually be set by radiation damage effects, which have been extensively studied in high-voltage electron microscopy. Using the appropriate cross-sections for the knock-on ballistic atomic displacement process in silicon \cite{Vanhellemont1993} and realistic beam current density of 0.1 A/m$^2$ (1 pC at 1 kHz repetition rate into a 100 micron diameter spot) we estimate 1\% atomic displacements to accumulate from 24/7 continuous operating exposure after 277 days. The key results of this first set of experiments are that the photoinjector operations, stability, and beam quality are sufficient and achievable in the presence of space charge effects for the nanopatterning technique, and that we understand the diffraction physics well enough to predict nanopatterning performance.

Following the UED experiments we collaborated with Norcada to design and fabricate a Si grating with 400 nm period in order to produce the nanopatterned beam.  The grating device is a 200 nm thick $500 \times 500$ $\mu$m Si membrane  located at the center of a 525 $\mu$m thick polygon chip of 3 mm diameter to fit in a standard TEM holder. At the center of the membrane, a $100 \times 100$ $\mu$m square area (Figure \ref{Fig04}a) contains a cut-through grating pattern with 200 nm wide peak and 200 nm valley (400 nm pitch). Those cuts have been aligned to the edge of the wafer and to the crystal plane. The grating experiment was performed in a 300 keV Titan transmission electron microscope due to the lack of imaging optics at the UED facility. At the orientation maximizing the $(2\overline{2}0)$ Bragg reflection, an aperture was introduced to block all but the direct beam, producing the bright field image shown in Figure \ref{Fig04}b.  We find that 80\% of the beam that passes through the thicker Si strips is deflected out of the zero-order spot, matching the results from the UED facility, and producing a strong 4:1 contrast ratio in the image plane when other spots are blocked by the aperture (Figure \ref{Fig04}c).

\begin{figure}[!htb]
	\centering
	\includegraphics[width=8cm]{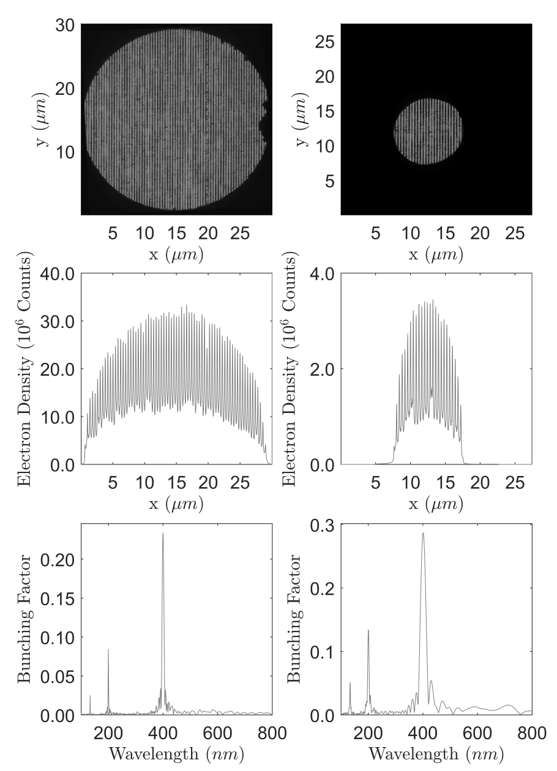}
	\caption{Top row shows bright field images of modulated electron beam for two different beam sizes illuminating different numbers of grating periods.  Middle row shows projected distributions, and bottom row shows FEL bunching factor.  On left the larger beam covers 69 periods of the grating structure, and on right the small beam covers 25 periods.  The smaller beam with fewer periods creates a narrower bunch with broader bandwidth.  Following emittance exchange these properties would be transferred to the time-frequency domain illustrating the simple and deterministic control of FEL pulse length and bandwidth possible with this method.}
	\label{Fig05}
\end{figure} 

Figure \ref{Fig04}d shows the amplitude of the Fourier transform of this distribution, known as the FEL bunching factor \cite{Bonifacio1990} (albeit measured in the transverse direction), showing a peak value near 0.5, similar to the maximum bunching factor seen after FEL saturation.  Figure \ref{Fig04}d also shows significant harmonic content that may be exploited to reach even shorter wavelengths than the fundamental.  The width of the spike is within a few percent of the transform limit, demonstrating the power of this technique to produce fully coherent bunching.  

Figure \ref{Fig05} demonstrates the ability to control the output beam properties by simple manipulation of the electron spot size at the Si grating.  In this case we have changed the electron spot size to vary the number of periods illuminated.  In this example the larger spot on the left results in 69 periods and the smaller spot on the right results in 25 periods of modulation.  The bandwidth of these pulses is 1.8\% and 5.4\% respectively, close to the transform limits of 1.5\% and 4.0\%. The number of nanobunches also determines the output pulse length.  For the images shown in Figure \ref{Fig05}, demagnification by quadrupole lenses and implementation of emittance exchange (which provides another factor of 6 demagnification \cite{Nanni2018}) these pulses would be just 0.1 fs and 0.3 fs long at a period of 1.24 nm.  The lower bunching factor ($\sim0.3$) seen here is an artifact of the microscope optics when using low TEM magnification.  At low magnification the objective lens is switched off moving the diffraction plane slightly out of the selection aperture location, so that the contrast is lower.

Diffraction from the silicon structure allows for the precise tailoring of the electron bunch pattern that eventually drives the x-ray phase. This tailoring will be transferred to the longitudinal dimension resulting in phase-controlled x-ray output pulses.  Relative slippage of one x-ray wavelength per undulator period occurs between x-rays and electrons during emission, limiting the ability to perform instantaneous phase shifts.  Nevertheless the x-ray properties can be manipulated in novel ways.  We can coherently control the frequency, bandwidth, pulse length and amplitude of the x-ray pulses. This method will thus provide tunable coherent attosecond to femtosecond photon beams with unmatched phase control. In addition, the beam patterning method allows nearly any desired time structure to be imposed on the beam by masking with semiconductor lithography. With distance across the mask corresponding, via emittance exchange, to time delay on an attosecond scale, it becomes possible to generate an initial time=0 pulse for x-ray pump, x-ray probe experiments with near-zero jitter, and to produce chirped beams at will, with full temporal coherence and deterministic pulse profiles, unlike the amplified noise on which SASE XFELs are based. 

The ability to produce large ($\sim$10 eV) bandwidths and application of chirped pulse compression will enable hard x-ray pulses with attosecond duration. An example of the novel capabilities enabled include probing the first step of chemical reactions, i.e. electronic rearrangements and charge migration, which in turn trigger the motion of nuclei and produce changes in the molecular skeleton. Attosecond x-ray pulses will help to address many open questions on the role of electronic correlations and electron-nuclear couplings in the initial phase of photochemical reactions. These studies are key to understanding the basic steps of life-giving processes such as photosynthesis, vision, and catalysis, both for industrial chemical production and for the enzymes which control human biochemistry. 

A compact source using a laser undulator based on these methods is limited in peak flux compared to large XFELs.  We estimate single shot flux at about 30 nJ limited by the low electron beam charge and energy. The compact source can provide a coherent seed much larger than SASE startup power to large XFELs for experiments that demand mJ pulses with the stability and flexibility that phase control allows.  The seed must have a power significantly greater than the SASE startup noise of $\sim10$ kW and high stability from shot to shot.  Such a seed would then transfer the deterministic phase properties to the higher power beam of a large XFEL.

In summary, we studied electron diffraction through thin single-crystal membranes both with and without lithographic patterning.  These studies include production of transversely modulated electron beams as the first step toward full control of the phase of XFEL output, which is expected to have a large impact on ultrafast x-ray science.  The predicted diffraction intensities closely match simulations and show nearly complete extinction of the beam in response to the lithographic patterning, as desired.  The resulting modulated beam has strong Fourier components at the modulation period, which may be tuned by varying the magnification with straightforward quadrupole optics.  We demonstrated that the electron beam produced by a photoinjector has the beam properties of energy spread and emittance needed to cleanly diffract through the membrane.

We gratefully acknowledge support from NSF Accelerator Science award 1632780 and NSF BioXFEL STC award 1231306. The UED work was performed at the SLAC MeV UED facility, which is supported in part by the DOE BES SUF Division Accelerator \& Detector R\&D program, the Linac Coherent Light Source (LCLS) Facility, and SLAC under contracts DE-AC02-05-CH11231 and DE-AC02-76SF00515. We acknowledge the use of facilities in the John M. Cowley Center for High Resolution Electron Microscopy at Arizona State University. The use of the Titan ETEM was supported in part by NSF Grant CBET-1604971.

\bibliography{sources03-12-2019.bib}

\end{document}